\documentclass[epj]{svjour}
\usepackage{graphics}
\usepackage{amssymb}
\usepackage{amsmath}
\usepackage{bm}
\begin{document}
\title{Isospin Dependence of Nucleon Correlations in Ground-State Nuclei}
\author{R. J. Charity\inst{1} \and W. H. Dickhoff\inst{2} \and L. G. Sobotka\inst{1,2} \and S. J. Waldecker\inst{3}
}                     
%
%
\institute{Department of Chemistry, Washington University, St. Louis, Missouri 63130, USA \and Department of Physics, Washington University, St. Louis, Missouri 63130, USA \and Department of Physics, University of Tennessee, Chattanooga, Tennessee 37403, USA}
\date{Received: date / Revised version: date}
%
\abstract{
The dispersive optical model (DOM) as presently implemented can investigate the isospin (nucleon asymmetry) dependence of the Hartree-Fock-like potential relevant for nucleons near the Fermi energy.
Data constraints indicate that a Lane-type potential adequately describes its asymmetry dependence.
Correlations beyond the mean-field can also be described in this framework, but this requires an extension that treats the non-locality of the Hartree-Fock-like potential properly. 
The DOM has therefore been extended to properly describe ground-state properties of nuclei as a function of nucleon asymmetry in addition to standard ingredients like elastic nucleon scattering data and level structure. 
Predictions of nucleon correlations at larger nucleon asymmetries can then be made after data at smaller asymmetries constrain the potentials that represent the nucleon self-energy.
A simple extrapolation for Sn isotopes generates predictions for increasing correlations of minority protons with increasing neutron number. Such predictions can be investigated by performing experiments with exotic beams.
The predicted neutron properties for the double closed-shell ${}^{132}$Sn nucleus exhibit similar correlations as those in ${}^{208}$Pb.
Future relevance of these studies for understanding the properties of all nucleons, including those with high momentum, and the role of three-body forces in nuclei are briefly discussed.
Such an implementation will require a proper treatment of the non-locality of the imaginary part of the potentials and a description of high-momentum nucleons as experimentally constrained by the $(e,e'p)$ reactions performed at Jefferson Lab.  
} 
\maketitle
\section{Introduction}
\label{intro}
The motion of nucleons inside of a nucleus, to a first approximation, 
can be described by a non-local mean-field potential. The nucleon self-energy provides a complete description of single-particle motion and can be represented as a complex, energy-dependent, non-local potential where the imaginary part encodes information on correlations beyond the mean-field contribution. 
The propagation of a nucleon including these correlations can be described by 
the single-particle propagator  $G_{\ell,j}(r,r';E)$ in an angular momentum basis. The propagator 
can be obtained from the self-energy by solving the Dyson equation~\cite{Dickhoff08}.
  
Elastic scattering angular distributions and reaction cross sections have a 
long history in which they are fitted using  complex optical-model potentials 
where the imaginary part describes the absorption from the elastic channel.
As the nucleon optical-model potential and the self-energy are both complex, it is not surprising that they are related. 
Mahaux and Sartor~\cite{Mahaux91} were the 
first to consider using such complex potentials to describe both reaction data 
such as elastic scattering and bound-state information at the same time. 
To achieve this, two important issues must be considered. 

First, the nucleon self-energy is non-local, 
but optical-model potentials are generally taken to be local. Historically, 
due to the more limited computational abilities in the past, 
calculations with non-local potentials were often impractical to perform within a 
reasonable time period, 
so most calculations were made with ``faster'' local potentials.  
However, it was 
recognized that the energy dependence of the depth of the real potential 
obtained in global optical-model fits, was not a real energy dependence, 
but a consequence of disregarding non-locality~\cite{Perey62}.

Secondly, the self-energy should obey  the relevant dispersion relations between 
its real and imaginary parts, a consequence of causality. 
Most standard optical potentials ignore this relationship~\cite{Becchetti69,Varner91,Koning03}. In order to remedy 
this situation and to allow a successful extrapolation of optical-models potentials, 
which are fitted to positive-energy data, down to the negative-energy regions 
and confront bound-state measurements, Mahaux and Sartor~\cite{Mahaux91} 
developed the dispersive optical model (DOM).  In this description, the real 
part of the nucleon self-energy can be decomposed into an energy-independent 
non-local part and an energy-dependent contribution
\begin{equation}
\text{Re}~\Sigma \left( \bm{r},\bm{r}^{\prime };E\right) =\text{Re}~\Sigma
\left( \bm{r},\bm{r}^{\prime };\varepsilon_{F}\right) +\Delta \mathcal{V}(\bm{r},\bm{r}%
^{\prime };E),  \label{eq:self}
\end{equation}%
where $\varepsilon_F$ is the Fermi energy, defined according to
\begin{eqnarray}
\label{Eq:Fermi}
\varepsilon_{F} &=& \frac{\varepsilon_{F}^{+} + \varepsilon_{F}^{-}}{2}\\
\varepsilon_{F}^{+} & = & M_{A+1} - (M_{A} + m) 
\label{Eq:Fermi+} \\
\varepsilon_{F}^{-} & = & M_{A} - (M_{A-1} + m) ,
\label{Eq:Fermi-}
\end{eqnarray}
where $\varepsilon_{F}^{+}$ and $\varepsilon_{F}^{-}$ represent the binding energy for adding or 
removing a nucleon, or alternatively, the single-particle energies of the 
valence particle and hole states.
The second term in Eq.~(\ref{eq:self}) represents the dispersive correction which 
can be determined from the imaginary component through the subtracted 
dispersion relation
\begin{eqnarray}
\lefteqn{\Delta \mathcal{V}(\bm{r},\bm{r}^{\prime };E)=}  \label{eq:sdr} \\
&&+\frac{1}{\pi }\mathcal{P}\int \text{Im}~\Sigma \left( \bm{r},\bm{r}^{\prime
};E^{\prime }\right) \left( \frac{1}{E^{\prime }-E}-\frac{1}{E^{\prime
}-\varepsilon_{F}}\right) dE^{\prime }.  \notag
\end{eqnarray}%

The non-local energy-independent term $\text{Re}\Sigma(\bm{r},\bm{r}';\varepsilon_F)$ can be 
approximated by a local energy-dependent term, and also the use of this 
local approximation requires the use of an imaginary potential which is a 
scaled version of the imaginary self-energy~\cite{Mahaux91}. 
These real and imaginary optical-model potentials are parameterized based on past experience in 
fitting elastic-scattering data and expectations from nuclear-matter 
calculations.

By fitting data in a sequence of isotopes (or isotones) it is possible to extract the isospin dependence of the DOM potentials thereby allowing predictions for nuclei further off stability and closer to the respective drip lines~\cite{Charity06,Charity07}.
In the most recent investigation~\cite{Mueller11} we have made an extended study with the DOM for 
spherical nuclei using known elastic nucleon scattering angular distributions and analyzing 
powers, reaction and total cross sections (the latter for neutrons only). 
At negative energies, we have included single-particle energies and 
spectroscopic factors extracted from ($e,e'p$) data when available.    
The DOM fits were obtained from proton and neutron measurements for 
Ca, Ni, Fe, Cr, Zr, Mo, Sn, and Pb isotopes.  See Ref.~\cite{Mueller11} 
for details of the parameterizations and the fits. 
From these fits we can extract information of the symmetry energy and the 
asymmetry dependence of effective masses and nucleon correlations. 

\section{Extracted potential and 
asymmetry dependence}
\subsection{Symmetry Energy}
\label{sec:symmetry energy}
\begin{figure}
\resizebox{0.5\textwidth}{!}{%
  \includegraphics{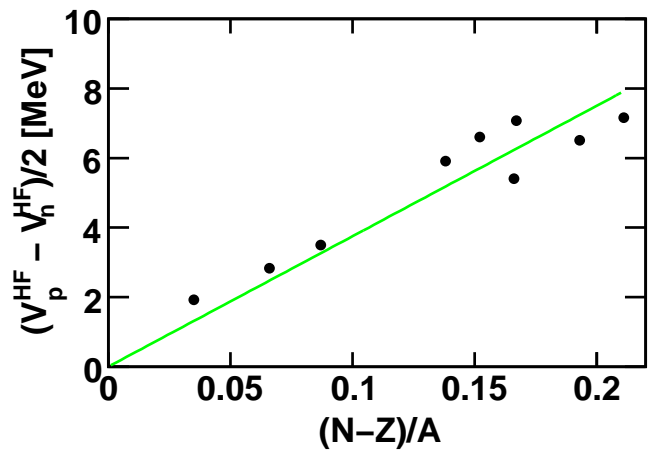}
}
\caption{Extracted differences in the depths of the proton and neutron real 
potentials as a function of symmetry.}
\label{fig:symmetry}       
\end{figure}

The symmetry energy is composed of a potential part and a 
kinetic-energy component. The potential part can be extracted from the 
$(N-Z)/A$
dependence of the real part of the fitted optical-model potential. 
This may be complicated by the  energy dependence 
(Sec.~\ref{sec:nonlocality}), which itself can be dependent on  $(N-Z)/A$.
In the DOM all potentials are defined relative to the Fermi energy, and 
thus we will look for the $(N-Z)/A$ dependence at this energy. 
The major part of the real potential was assumed to have a Wood-Saxon 
form factor and its 
depth is constrained by reproducing the experimental Fermi energy while 
fitting the experimental data. Following the Lane model~\cite{Lane62} 
we expect this depth to have isoscalar and isovector components
\begin{equation}
  V = V_{0} \pm V_{1} \frac{N-Z}{A} ,
\end{equation} 
where the plus sign refers to protons and the minus sign refers to neutrons. For nuclei where both the 
proton and neutron depths are extracted, we would expect the following 
relationship
\begin{equation}
\frac{V_{p}-V_{n}}{2} = V_{1} \frac{N-Z}{A} .
\end{equation}
The differences in the extracted potentials from \cite{Mueller11} is plotted 
against $(N-A)/A$ in 
Fig.~\ref{fig:symmetry} and expected linear relationship is confirmed.
From a fit to these data, the potential part of the symmetry energy   
 is found to be $V_{1}$=18.4~MeV. This can be compared to the value of 
13.1~MeV extracted by Varner et al. \cite{Varner91} in their 
global-optical-model fits.

When our value of $V_{1}$ is added to the standard value of the 
symmetry kinetic energy for saturated nuclear matter of  12 MeV \cite{Eisenberg87},
 a total symmetry energy of 30-31~MeV is obtained. This can be compared to 
those extracted from the Seeger's mass formula~\cite{Seeger68} or the 
droplet model~\cite{Myers77} with values of 30.6 and 36.8~MeV, respectively.
Alternatively it is also comparable to the value of 32.4 MeV obtained from combined constraints of a global mass fit and the mass differences between isobaric analog states~\cite{Danielewicz09}. 

\subsection{Asymmetry dependence of non-locality}
\label{sec:nonlocality}
It has been long known that to fit elastic-scattering data and total 
reaction cross sections, the depth of the local real potential must exhibit a specific
energy-dependence. 
However, a substantial part of this energy-dependence mimics the 
effect of a static non-local potential $V(\bm{r},\bm{r}')$, as shown by Perey and Buck~\cite{Perey62}. 
The magnitude of the non-locality can be characterized by the momentum-dependent effective mass $\widetilde{m}$ which can be obtained from the energy-dependence of the real potential
\begin{equation}
\frac{\widetilde{m}}{m} = 1 - \frac{dV(E,r)}{dE}
\end{equation}
where $m$ is the nucleon rest mass. 
The isoscalar ($V_{0}$) and isovector ($V_{1}$) 
components can have different non-localities and 
thus different energy dependences that we have allowed for in our 
dispersive-optical-model fits. The isoscalar dependence is much better 
defined and the extracted values in the center of the nucleus increase from $\widetilde{m}=0.49$ $m$ in Ca nuclei to 0.62 $m$ in Pb nuclei. 
The extracted isovector non-locality is much smaller than the isoscalar value, a result consistent with Rook~\cite{Rook74}. From the extracted value, we deduce a small, but still statistically significant, difference in proton and neutron effective masses 
in the center of the $^{208}$Pb nucleus of $\widetilde{m}_{p}-\widetilde{m}_{n} = 0.054$ $m$.  

\subsection{Asymmetry dependence of correlations}
The imaginary potential gives information on the correlations of nucleons 
in the nucleus and is directly related to the absorption from the elastic 
channel. 
From a long history of  optical-model fits, it has been determined that at 
low bombarding energies the absorption is mostly confined to the surface of the nucleus 
while at high energies the absorption is over the whole volume of the nucleus. 
Thus in many global optical-model studies one assumes a surface-type 
imaginary potential with a radial form factor given by the derivative 
of a Wood-Saxon and a volume-type imaginary potential  where the form factor 
is given by a Wood-Saxon. The change over from surface to volume absorption 
occurs at an energy of around 50~MeV but is quite gradual. 

The surface potential is confined to energies not too far from the 
Fermi energy and can be largely associated with long-range correlations, \textit{i.e.},
the coupling of the nucleons to surface vibrations and giant resonances.
On the other hand, the volume potential is restricted to energies further 
away from the Fermi energy and is dominated by tensor and short-range 
correlations. 

\begin{figure}
\resizebox{0.48\textwidth}{!}{%
  \includegraphics{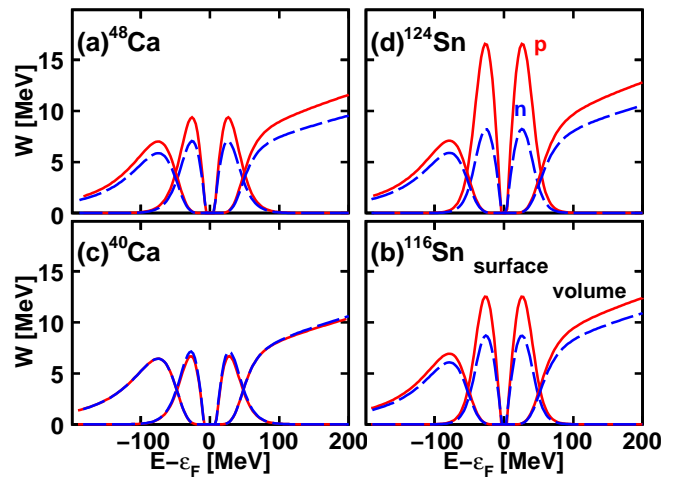}
}
\caption{Extracted energy dependence of the imaginary surface and 
volume potentials for both protons and neutrons for the indicated systems.}
\label{fig:imag}       
\end{figure}

An example of the fitted energy dependence of the magnitudes 
of the surface and volume potentials is displayed in Fig.~\ref{fig:imag}.  
To obtain the dispersion correction and to extract correlations, 
the imaginary potential must be defined at negative as well as
positive energies. To understand the energy dependences in 
Fig.~\ref{fig:imag}, the following logic was employed 
parameterizing these potentials in the fits. From simple 
considerations, the total imaginary potential 
should be zero at the Fermi energy~\cite{Mueller11} as can also be inferred directly from Eq.~(\ref{eq:sdr}). 
Following Mahaux and Sartor, the 
energy-dependence of the surface potential is assumed symmetric around the 
Fermi energy~\cite{Mahaux91}. This symmetry is imposed on the volume component
at  energies not too far from the Fermi energy, but at larger energies, we employ the 
deviation from symmetry that was used by Mahaux~\cite{Mahaux91}. 
It is based on nuclear matter calculations and associated with the much larger phase space associated with two-particle--one-hole as compared to one-particle--two-hole excitations which govern the size of the imaginary potential above and below the Fermi energy, respectively.

The fitted potentials exhibit the following behavior. First, the proton 
and neutron potentials are essentially equal for the one $N$=$Z$ case 
[$^{40}$Ca, Fig.~\ref{fig:imag}(c)] 
considered. Otherwise for the  $N>Z$ cases, the proton imaginary 
potential is larger than the neutron potential. The larger proton 
imaginary potentials imply that protons experience larger correlations in these 
$N>Z$ nuclei than the majority neutrons. Note,  there are no $N<Z$ data to check 
if the expected opposite trend holds. 

\begin{figure}
\resizebox{0.48\textwidth}{!}{%
  \includegraphics{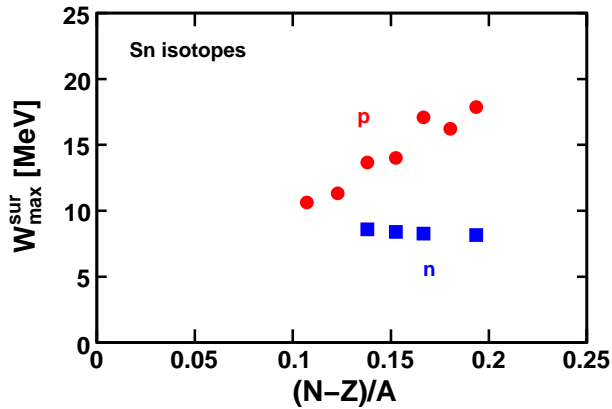}
}
\caption{Extracted maximum value of the imaginary surface potential for 
both proton and neutrons on Sn isotopes plotted as a function of asymmetry.}
\label{fig:sn}       
\end{figure}

Second, the asymmetry dependence is quite strong for the proton surface 
component; this is especially clear for the Sn isotopes. 
To further illustrate this, the extracted peak value of the surface imaginary 
potential is plotted against asymmetry in Fig.~\ref{fig:sn} for Sn isotopes. 
Protons display a linearly increasing dependence on asymmetry, while for neutrons this dependence is almost 
non-existent. The dependence observed for protons in lighter nuclei is somewhat 
more complicated possibly due to the presence of multiple closed shells 
but a general increase with asymmetry is still observed and  
 neutrons continue to show an insensitivity to this quantity. These results 
imply that protons experience increasing long-range correlations in more neutron-rich 
systems while neutron correlations do not change significantly.
   
In the independent-particle model (IPM), the strength of a single-particle level 
is located at a single energy. However, the action of the correlations 
spreads this strength out to higher and lower energies, and the energy 
distribution is called the spectral function. In order to obtain realistic 
spectral functions it is necessary to use non-local potentials. 
The real energy-dependent optical-model potential not due to the dispersive contribution is then replaced by an 
``equivalent'' non-local Hartree-Fock-like (HF) potential and the imaginary potentials are scaled  
to remove the effects of non-locality~\cite{Dickhoff10}. The resulting potential can then be considered as the self-energy of the nucleon 
and inserted into the Dyson equation to solve for the single-particle propagator $G_{\ell,j}(r,r';E)$. The spectral function is then
given by 
\begin{equation}
S_{\ell,j}(E) =  \frac{1}{\pi} \int dr\ r^2\ \textrm{Im}\ G_{\ell,j}(r,r;E) .
\end{equation}
For a valence level, this strength consists of a delta function at the 
IPM energy and continuum contributions at higher and lower energies. 
The spectroscopic factor represents the integral of this delta-function 
component and characterizes the 
reduction of the localized strength at the IPM energy due to correlations.
Such reductions have been clearly demonstrated on the basis of the analysis of the $(e,e'p)$ reaction~\cite{Lapikas93}. 
We further note that by implementing the non-local HF potential it is no longer necessary to make approximations for the calculation of spectral functions and occupation numbers~\cite{Mueller11} as in the original version of the DOM~\cite{Mahaux91}.

\begin{figure}[t]
\resizebox{0.48\textwidth}{!}{%
  \includegraphics{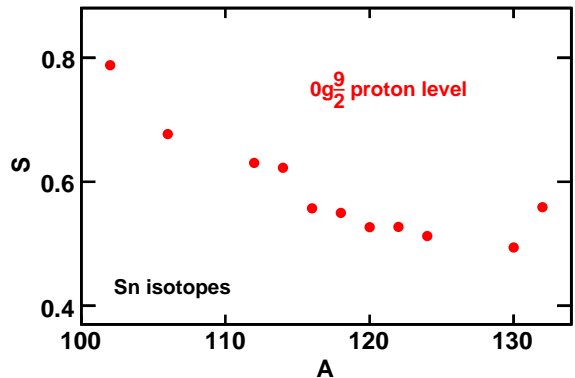}
}
\caption{Spectroscopic factors relative to the independent-particle-model values deduced for the $0g_{9/2}$ proton levels in Sn isotopes from 
the DOM analysis.}
\label{fig:SFsn}       
\end{figure}

\begin{figure}[b]
\resizebox{0.48\textwidth}{!}{%
  \includegraphics{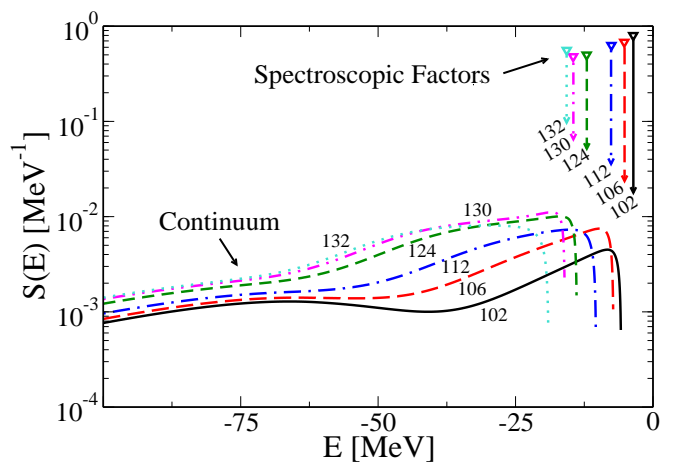}
}
\caption{ Strength functions of the $0g_{9/2}$ proton  orbit 
in different Sn isotopes obtained with the non-local calculations.
The curves represent the continuum contribution of the strength function 
and are labeled by the appropriate mass number. Also indicated is the 
location of the $0g_{9/2}$ quasihole level in the different isotopes. 
The height of the corresponding vertical lines identifies the spectroscopic 
factor for each isotope.
} 
\label{fig:sfg92}
\end{figure}

Examples of spectroscopic factors and spectral strength functions are 
shown in Figs.~\ref{fig:SFsn} and \ref{fig:sfg92}, respectively, for the 
valence $0g_{9/2}$ proton hole states in Sn isotopes.  We have used a linear
 extrapolation of the asymmetry dependence of the surface imaginary potential in 
Fig.~\ref{fig:sn} to obtain predictions for $^{102}$Sn, $^{106}$Sn, $^{130}$Sn, 
and $^{132}$Sn also shown in these figures.  
The increasing surface imaginary potential with asymmetry 
is associated with increasing 
proton correlations. This gives rise to reductions in the spectroscopic factor 
(the local strength at the IPM energy)  with an accompanying increase of the strength in the continuum as illustrated 
in the extracted spectral functions of Fig.~\ref{fig:sfg92}. 
The predicted spectroscopic factors in Fig.~\ref{fig:SFsn} 
decrease by $\sim$30\% from $^{102}$Sn to $^{130}$Sn. The subsequent 
small rise in spectroscopic factor for $^{132}$Sn is associated with the 
closure of the neutron shell.  
As most of the asymmetry dependence is due to the surface-imaginary component, the changes in the spectral 
functions are mostly due to increased long-range correlations.
The possible importance of low-energy tensor correlations to explain this behavior was suggested by the \textit{ab initio} analysis of the Faddeev random phase approximation (FRPA) self-energy for different Ca isotopes~\cite{Waldecker2011}.
The non-local HF potential used for the Sn isotopes was subjected to the additional constraint of fitting the charge radii of ${}^{112}$Sn and ${}^{124}$Sn~\cite{deVries1987}.

\begin{figure}
\begin{center}
\resizebox{0.35\textwidth}{!}{%
  \includegraphics{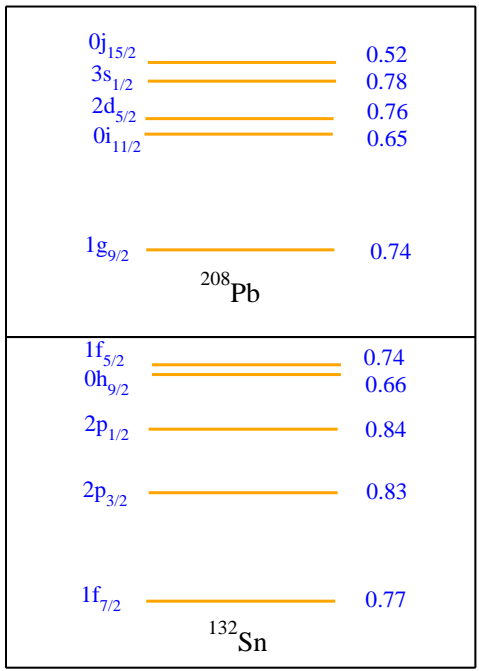}
}
\caption{ Properties of neutron states above the ${}^{208}$Pb core in the top panel compared with those above the 
${}^{132}$Sn core. In each panel, the levels are
labeled on the left and the corresponding spectroscopic factors are given on
the right. The levels and numbers are from the DOM implementation with a non-local HF potential~\protect{\cite{Dickhoff10}}.
} 
\label{fig:Levels132-208}
\end{center}
\end{figure}

Based on the analysis of elastic-scattering data the neutron properties appear to change considerably less than those of protons with increasing neutron number. 
Extrapolating the neutron potentials to ${}^{132}$Sn, using the non-local HF potential generates neutron particle states and related spectroscopic factors that are displayed in Fig.~\ref{fig:Levels132-208}.
A comparison with the corresponding neutron states above the ${}^{208}$Pb core demonstrates that the spectroscopic factors are of very similar nature confirming the double-closed-shell character of this exotic nucleus.

Neutron transfer data on ${}^{132}$Sn in inverse kinematics generate unrealistic results when the traditional analysis of these experiments is performed, yielding unphysical spectroscopic factors that can be larger than 1~\cite{Jones2010,Jones2011} when normalizing these quantities as probabilities~\cite{Dickhoff08}. 
While the present analysis of these reactions yields useful information about the relative strength of these single-particle states and allows a comparison between ${}^{132}$Sn and ${}^{208}$Pb~\cite{Jones2010}, it is much more preferable to analyze these reactions in a way that generates physically reasonable outcomes.
A step in this direction was recently made in Ref.~\cite{Nguyen2011}.
The description of the ${}^{132}$Sn($d,p$) transfer reaction was based on the adiabatic distorted wave approximation (ADWA)~\cite{Johnson05} and included traditional optical potentials as well as DOM ingredients.
The latter also generates the relevant overlap function for the neutron transfer.
The results of Ref.~\cite{Nguyen2011} indicate that the shape of the transfer cross sections including the one for ${}^{132}$Sn can be well described by the DOM potentials.
The DOM-generated cross sections are however typically larger as compared to those from traditional optical potentials. 
When suitably extrapolated for ${}^{132}$Sn they require therefore a larger reduction factor to fit the experimental cross section. 
In turn, this allows the reduction factor to be properly interpreted as a reasonable estimate of the spectroscopic factor.
With these ingredients the $f_{7/2}$ neutron spectroscopic factor deduced from the data becomes 0.72 compared to the 0.77 obtained directly from the DOM potential and displayed in Fig.~\ref{fig:Levels132-208}. 
This satisfactory outcome is also obtained for the analysis of ${}^{48}$Ca but this description fails to generate useful results for ${}^{208}$Pb which may be due to the need for treating target excitation in that nucleus since the conventional analysis is similarly inconsistent~\cite{Nguyen2011}.
It is noteworthy that the Ca and Sn results generate similar spectroscopic factors as those obtained from the $(e,e'p)$ reaction for double-closed-shell nuclei near stability~\cite{Lapikas93}.

Further extrapolations towards the expected drip line for neutrons in Sn isotopes yields an even more dramatic reduction of the proton spectroscopic factors.
Corresponding small reduction factors have also been extracted from heavy-ion knockout reactions for minority nucleons while the majority species appears to require almost no reduction~\cite{Gade04,Gade08}.
Transfer reactions appear to suggest a considerably smaller dependence of correlations on nucleon asymmetry~\cite{Lee10,Flavigny13} which is also suggested by FRPA \textit{ab initio} calculations~\cite{BaDi09}.
\begin{figure}
\begin{center}
\resizebox{0.48\textwidth}{!}{
  \includegraphics{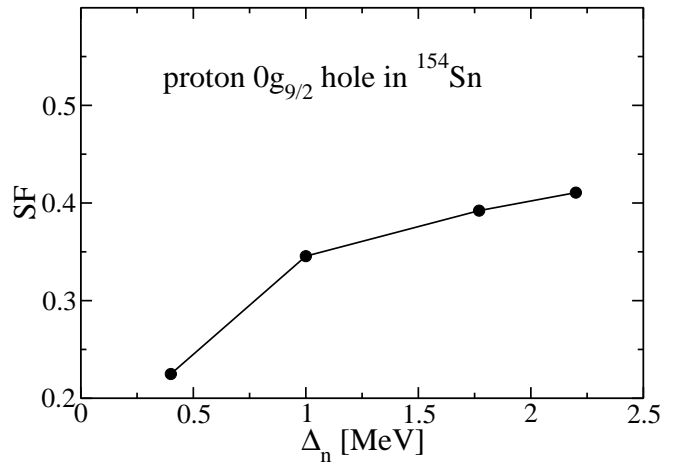}
}
\caption{ Spectroscopic factor of the proton $0g_{9/2}$ hole state of ${}^{154}$Sn as function of $\Delta_n$, the neutron separation energy.}
\label{fig:154Snsf}
\end{center}
\end{figure}
We illustrate the DOM extrapolation in Fig.~\ref{fig:154Snsf} for ${}^{154}$Sn~\cite{Moller95},  a possible candidate for the neutron drip line.
The proton $0g_{9/2}$ spectroscopic factor is shown as a function of the unknown separation energy of the last neutron.
It is clear from the figure that the proximity of the neutron continuum has important consequences for the strength of correlations as measured by the valence proton spectroscopic factor.
The sensitivity of the reduction of the spectroscopic factor demonstrates the important role that the continuum could play in determining the size of the valence spectroscopic factor.
This feature was also pointed out in Ref.~\cite{Jensen11} on the basis of an \textit{ab initio} coupled-cluster calculation with a proper treatment of the continuum.
While such calculations point to a sizable reduction, they fail to adequately account for the reduction in stable closed-shell nuclei such as ${}^{16}$O~\cite{Leuschner94}.
The main effect of reducing the strength of the valence hole due to the presence of the continuum appears to be related to the stronger low-energy surface absorption which implies that the lost strength resides in the nearby continuum in the way it is illustrated in Fig.~\ref{fig:sfg92}. 
This also implies that the occupation number of such a valence orbit is much less sensitive to the proximity of the continuum~\cite{Mueller11}.

While the DOM is capable of generating strong reductions of the spectroscopic strength in principle, which are yet to be conclusively established experimentally, it is not yet capable of generating the large differences between the minority and majority species suggested by heavy-ion knockout reactions but not confirmed by corresponding transfer reactions.
A recent knockout experiment on ${}^{36}$Ca suggests reduction factors of 0.75 and 0.22 for the majority proton $d_{3/2}$ and minority neutron $s_{1/2}$ orbits, respectively~\cite{Shane12}.
DOM extrapolations for $^{36}$Ca generate 0.70 and 0.65 for the proton and neutron orbit, respectively.
It is useful to note however that the assumed energy dependence of the imaginary part of the DOM potentials near the Fermi energy does not yet reflect the more realistic pole structure that is present in the nucleon self-energy and also exhibits no state dependence which can be quite important near the Fermi energy~\cite{Waldecker2011}. 

\section{Future Developments}
 
The introduction of the non-local HF potential and the attendant renormalization of the imaginary potentials make it possible to interpret the DOM potential as representing the nucleon self-energy.
The solution of the Dyson equation then generates, for energies below the Fermi energy, the ingredients which allow the calculation of the expectation value in the ground state of all one-body operators as well as the contribution to the ground-state energy from two-body interactions~\cite{Dickhoff08}.
This opens up the possibility to utilize accurately determined charge densities as additional ingredients in the DOM fits.
In addition, the calculated ground-state energy from the DOM propagator can provide information on the empirical contribution of three-body forces to nuclear binding.
Initial calculations of these quantities were performed in Ref.~\cite{Dickhoff10} showing that it was quite feasible to describe the radius of the nuclear charge distribution of ${}^{40}$Ca but not its detailed radial dependence.
The energy of the ground state was also substantially underestimated by this initial calculation.
Another serious flaw was the inability to reproduce the number of particles from the DOM spectral functions, typically overestimating this quantity by more than 10\%.

Recent \textit{ab initio} calculations of the nucleon self-energy have clarified some of these issues and indicated how they can be addressed~\cite{Waldecker2011,Dussan11}.
Both studies clarify that the calculated imaginary parts of the self-energy exhibit a very substantial angular momentum dependence.
This state dependence cannot be represented by the local implementation of the DOM potentials that consequently leads among others to an overestimate of the particle number since the absorption below the Fermi energy is independent of angular momentum.
While non-locality appears inescapable from an \textit{ab initio} perspective, it appears to be possible to represent it in a rather standard manner~\cite{Dussan11,Waldecker2011}, \textit{i.e.} by a Gaussian form as originally suggested in Ref.~\cite{Perey62}.
Also the symmetry assumption of the absorption above and below the Fermi energy is called into question by the microscopic calculations.
DOM implementations now in progress~\cite{Mahzoon2013} therefore include explicitly non-local imaginary potentials as well as abandon the assumption of symmetric surface absorption above and below the Fermi energy.
These new functional forms appear to allow an accurate representation of the nuclear charge density as well as an adequate description of the high-momentum spectral functions generated by JLab experiments~\cite{Rohe04}.
The latter feature implies that a substantially better description of the ground-state energy can be expected as shown in Ref.~\cite{Muther95}.
Future DOM implementations will therefore be capable of assessing the empirical relevance of three-body contributions to nuclear binding.

As illustrated in the present paper, DOM potentials can easily be used to extrapolate to more exotic nuclei and therefore provide an excellent guide to predict future results at rare isotope facilities.
Such new experiments will then be able to clarify in more detail how protons and neutrons behave at even more extreme values of nucleon asymmetry.
The possibility in future to accurately describe the details of the nuclear charge distribution will widen the scope of the DOM and may also lead to further insights into the properties of neutrons in the ground states of exotic nuclei.
\section*{Acknowledgement}
This work was supported by the U.S. Department of Energy,
Division of Nuclear Physics under grant DE-FG02-87ER-40316 and the U.S.
National Science Foundation under grant PHY-0968941.
%
%
%
 \bibliographystyle{plain}
 \bibliography{level_deB}

\begin{thebibliography}{10}

\bibitem{BaDi09}
C.~Barbieri and W.~H. Dickhoff.
\newblock {\em Int. J. Mod. Phys. A}, 24:2060, 2009.

\bibitem{Becchetti69}
F.~D. {Becchetti Jr.} and G.~W. Greenlees.
\newblock {\em Phys. Rev.}, 182:1190, 1969.

\bibitem{Charity07}
R.~J. Charity, J.~M. Mueller, L.~G. Sobotka, and W.~H. Dickhoff.
\newblock {\em Phys. Rev. C}, 76(4):044314, 2007.

\bibitem{Charity06}
R.~J. Charity, L.~G. Sobotka, and W.~H. Dickhoff.
\newblock {\em Phys. Rev. Lett.}, 97:162503, 2006.

\bibitem{Danielewicz09}
P.~Danielewicz and J.~Lee.
\newblock {\em Nucl. Phys.}, A818:36, 2009.

\bibitem{deVries1987}
H.~{de Vries}, C.~W. {de Jager}, and C.~{de Vries}.
\newblock {\em At. Data Nucl. Data Tables}, 36:495, 1987.

\bibitem{Dickhoff08}
W.~H. Dickhoff and D.~{Van Neck}.
\newblock {\em Many-Body Theory Exposed!, 2nd edition}.
\newblock World Scientific, New Jersey, 2008.

\bibitem{Dickhoff10}
W.~H. Dickhoff, D.~Van~Neck, S.~J. Waldecker, R.~J. Charity, and L.~G. Sobotka.
\newblock {\em Phys. Rev. C}, 82(5):054306, 2010.

\bibitem{Dussan11}
H.~Dussan, S.~J. Waldecker, W.~H. Dickhoff, H.~M{\"u}ther, and A.~Polls.
\newblock {\em Phys. Rev. C}, 84:044319, 201.

\bibitem{Eisenberg87}
J.~M. Eisenberg and W.~Greiner.
\newblock {\em Nuclear Theory - Volume I}.
\newblock North-Holland, Amsterdam, 1987.

\bibitem{Flavigny13}
F.~Flavigny, A.~Gillibert, L.~Nalpas, A.~Obertelli, N.~Keeley, C.~Barbieri,
  D.~Beaumel, S.~Boissinot, G.~Burgunder, A.~Cipollone, A.~Corsi, J.~Gibelin,
  S.~Giron, J.~Guillot, F.~Hammache, V.~Lapoux, A.~Matta, E.~C. Pollacco,
  R.~Raabe, M.~Rejmund~N. {de Se«reville}, A.~Shrivastava, A.~Signoracci, and
  Y.~Utsuno.
\newblock {\em Phys. Rev. Lett.}, 110:122503, 2013.

\bibitem{Gade08}
A.~Gade, P.~Adrich, D.~Bazin, M.~D. Bowen, B.~A. Brown, C.~M. Campbell, J.~M.
  Cook, T.~Glasmacher, P.~G. Hansen, K.~Hosier, S.~McDaniel, D.~McGlinchery,
  A.~Obertelli, K.~Siwek, L.~A. Riley, J.~A. Tostevin, and D.~Weisshaar.
\newblock {\em Phys. Rev. C}, 77:044306, 2008.

\bibitem{Gade04}
A.~Gade, D.~Bazin, B.~A. Brown, C.~M. Campbell, J.~A. Church, D.~C. Dinca,
  J.~Enders, T.~Glasmacher, P.~G. Hansen, Z.~Hu, K.~W. Kemper, W.~F. Mueller,
  H.~Olliver, B.~C. Perry, L.~A. Riley, B.~T. Roeder, B.~M. Sherrill, J.~R.
  Terry, J.~A. Tostevin, and K.~L. Yurkewicz.
\newblock {\em Phys. Rev. Lett.}, 93:042501, 2004.

\bibitem{Jensen11}
{\O}.~Jensen, G.~Hagen, M.~{Hjorth-Jensen}, B.~A. Brown, and A.~Gade.
\newblock {\em Phys. Rev. Lett.}, 107:032501, 2011.

\bibitem{Johnson05}
R.~C. Johnson.
\newblock {\em AIP Conf. Proc.}, 791:128, 2005.

\bibitem{Jones2010}
K.~L. Jones et~al.
\newblock {\em Nature Letters}, 465:454, 2010.

\bibitem{Jones2011}
K.~L. Jones et~al.
\newblock {\em Phys. Rev. C}, 84:034601, 2011.

\bibitem{Koning03}
A.~J. Koning and J.~P. Delaroche.
\newblock {\em Nucl. Phys. A}, 713:231, 2003.

\bibitem{Lane62}
A.~M. Lane.
\newblock {\em Nucl. Phys.}, 35:676, 1962.

\bibitem{Lapikas93}
L.~Lapik{\'a}s.
\newblock {\em Nucl. Phys.}, A553:297c, 1993.

\bibitem{Lee10}
Jenny Lee, M.~B. Tsang, D.~Bazin, D.~Coupland, V.~Henzl, D.~Henzlova,
  M.~Kilburn, W.~G. Lynch, A.~M. Rogers, A.~Sanetullaev, A.~Signoracci, Z.~Y.
  Sun, M.~Youngs, K.~Y. Chae, R.~J. Charity, H.~K. Cheung, M.~Famiano,
  S.~Hudan, P.~O'Malley, W.~A. Peters, K.~Schmitt, D.~Shapira, and L.~G.
  Sobotka.
\newblock {\em Phys. Rev. Lett.}, 104:112701, 2010.

\bibitem{Leuschner94}
M.~Leuschner, J.~R. Calarco, F.~W. Hersman, E.~Jans, G.~J. Kramer,
  L.~Lapik{\'a}s, G.~{van der Steenhoven}, P.~K.~A. {de Witt Huberts}, H.~P.
  Blok, N.~Kalantar-Nayestanaki, and J.~Friedrich.
\newblock {\em Phys. Rev. C}, 49:955, 1994.

\bibitem{Mahaux91}
C.~Mahaux and R.~Sartor.
\newblock {\em Adv. Nucl. Phys.}, 20:1, 1991.

\bibitem{Mahzoon2013}
M.~H. Mahzoon, S.~J. Waldecker, R.~J. Charity, W.~H. Dickhoff, and H.~Dussan.
\newblock to be published, 2013.

\bibitem{Moller95}
P.~M\"{o}ller, J.~R. Nix, W.~D. Myers, and W.~J. Swiatecki.
\newblock {\em Atomic Data Nucl. Data Tables}, 59:185, 1995.

\bibitem{Mueller11}
J.~M. Mueller, R.~J. Charity, R.~Shane, L.~G. Sobotka, S.~J. Waldecker, W.~H.
  Dickhoff, A.~S. Crowell, J.~H. Esterline, B.~Fallin, C.~R. Howell,
  C.~Westerfeldt, M.~Youngs, B.~J. Crowe, and R.~S. Pedroni.
\newblock {\em Phys. Rev. C}, 83:064605, 2011.

\bibitem{Muther95}
H.~M{\"u}ther, A.~Polls, and W.~H. Dickhoff.
\newblock {\em Phys. Rev. C}, 51:3040, 1995.

\bibitem{Myers77}
W.~D. Myers.
\newblock {\em Droplet Model of Atomic Nuclei}.
\newblock Plenum, New York, 1977.

\bibitem{Nguyen2011}
N.~B. Nguyen, S.~J. Waldecker, F.~M. Nunes, R.~J. Charity, and W.~H. Dickhoff.
\newblock {\em Phys. Rev. C}, 84:044611, 2011.

\bibitem{Perey62}
F.~Perey and B.~Buck.
\newblock {\em Nucl. Phys.}, 32:353, 1962.

\bibitem{Rohe04}
D.~Rohe, C.~S. Armstrong, R.~Asaturyan, O.~K. Baker, S.~Bueltmann, C.~Carasco,
  D.~Day, R.~Ent, H.~C. Fenker, K.~Garrow, A.~Gasparian, P.~Gueye, M.~Hauger,
  A.~Honegger, J.~Jourdan, C.~E. Keppel, G.~Kubon, R.~Lindgren, A.~Lung, D.~J.
  Mack, J.~H. Mitchell, H.~Mkrtchyan, D.~Mocelj, K.~Normand, T.~Petitjean,
  O.~Rondon, E.~Segbefia, I.~Sick, S.~Stepanyan, L.~Tang, F.~Tiefenbacher,
  W.~F. Vulcan, G.~Warren, S.~A. Wood, L.~Yuan, M.~Zeier, H.~Zhu, and
  B.~Zihlmann.
\newblock {\em Phys. Rev. Lett.}, 93:182501, 2004.

\bibitem{Rook74}
R.~J. Rook.
\newblock {\em Nucl. Phys.}, A222:596, 1974.

\bibitem{Seeger68}
P.~A Seeger.
\newblock Technical Report LA-DC-8950a, Los Alamos Sci. Lab., 1968.

\bibitem{Shane12}
R.~Shane, R.~J. Charity, L.~G. Sobotka, D.~Bazin, B.~A. Brown, A.~Gade, G.~F.
  Grinyer, S.~McDaniel, A.~Ratkiewicz, D.~Weisshaar, A.~Bonaccorso, and J.~A.
  Tostevin.
\newblock {\em Phys. Rev. C}, 85:064612, 2012.

\bibitem{Varner91}
R.~L. {Varner}, W.~J. Thompson, T.~L. McAbee, E.~J. Ludwig, and T.~B. Clegg.
\newblock {\em Phys. Rep.}, 201:57, 1991.

\bibitem{Waldecker2011}
S.~J. Waldecker, C.~Barbieri, and W.~H. Dickhoff.
\newblock {\em Phys. Rev. C}, 84:034616, 2011.

\end{thebibliography}
%
%
%

\end{document}